\documentclass[
  aps,
  prl,
  reprint,
  preprintnumbers,
  showpacs,
  groupedaddress,
  amsmath,
  amssymb,
  floatfix]{revtex4-1}
\usepackage{graphicx,epsfig,dcolumn,multirow}
\usepackage{natbib}
\newcolumntype{d}[1]{D{.}{.}{#1}}

\RequirePackage{xspace}

\newcommand{\simgt}{\,\hbox{\lower0.6ex\hbox{$\sim$}\llap{\raise0.6ex\hbox{$>$}}}\,}
\newcommand{\simlt}{\,\hbox{\lower0.6ex\hbox{$\sim$}\llap{\raise0.6ex\hbox{$<$}}}\,} 

\newcommand{\ov}[1]{\bar{#1}}

\newcommand{\fig}[1]{Fig.~\ref{#1}}

\newcommand{\gev}{\ensuremath{\, \mathrm{GeV}}}

\newcommand{\Bcal}{\mathcal{B}}

\usepackage{pstricks}
\newcmykcolor{darkgreen}{1 0 0.6 0.5}  

\begin{document}

\preprint{TTP12-010, CERN-PH-TH-2012-101}

\title{
Joint analysis of Higgs decays and electroweak precision observables\\ 
in the Standard Model with a sequential fourth generation 
}

\author{Otto Eberhardt$^{\,a}$, Geoffrey Herbert$^{\,b}$, Heiko
  Lacker$^{\,b}$,\\ Alexander
  Lenz$^{\,c}$,  Andreas Menzel$^{\,b}$, Ulrich 
  Nierste$^{\,a}$, and Martin Wiebusch$^{\,a}$  
\vspace{0.6cm}}

\affiliation{
\mbox{$^{a}$ Institut f\"ur Theoretische Teilchenphysik,
Karlsruhe Institute of Technology, D-76128 Karlsruhe, Germany,}
\mbox{email: otto.eberhardt@kit.edu, ulrich.nierste@kit.edu, 
              martin.wiebusch@kit.edu}\\                
\mbox{$^{b}$ Humboldt-Universit\"at zu Berlin,
                   Institut f\"ur Physik,
                   Newtonstr. 15,
                   D-12489 Berlin, Germany,}\\
\mbox{e-mail: geoffrey.herbert@physik.hu-berlin.de, 
              lacker@physik.hu-berlin.de, amenzel@physik.hu-berlin.de}\\
\mbox{$^{c}$ CERN - Theory Divison, PH-TH, Case C01600, CH-1211 Geneva 23,
                {e-mail: alenz@cern.ch}}
}  
\date{\today}

\begin{abstract}
We analyse the impact of LHC and Tevatron Higgs data on the viability of the
Standard Model with a sequential fourth generation (SM4), assuming Dirac
neutrinos and a Higgs mass of 125 \gev. To this end we perform a combined fit to
the signal cross sections of $pp\to H\to \gamma\gamma,ZZ^*,WW^*$ at the LHC, to
$p\bar p\to VH\to Vb\bar b$ ($V=W,Z$) at the Tevatron and to the electroweak
precision observables. Fixing the mass of the fourth generation down-type quark
$b^\prime$ to 600$\,\gev$ we find best-fit values of $m_{t^\prime}=632\,\gev$,
$m_{l_4}=113.6\,\gev$ and $m_{\nu_4}=58.0\,\gev$ for the other fourth-generation
fermion masses.  We compare the $\chi^2$ values and pulls of the different
observables in the three and four-generation case and show that the data is
better described by the three-generation Standard Model. We also investigate the
effects of mixing between the third and fourth-generation quarks and of a future
increased lower bound on the fourth-generation charged lepton mass of
$250\;\text{GeV}$.
\end{abstract}


\maketitle
%
%
%
\section{Introduction}
%
%
While the Standard Model (SM) possesses a minimal boson field content,
it indulges itself in the luxury of replicated fermion generations.  It
is difficult to predict the number of generations from fundamental
theoretical principles; the determination of the correct number of
fermion families is ultimately an experimental task. A sequential
fourth generation is non-decoupling, meaning that its effect on certain
observables does not vanish in the limit of infinitely heavy
fourth-generation fermions.  Among these observables are the
gluon-fusion Higgs production cross section and the decay rate of $H\to
\gamma \gamma$.  This feature makes the SM with four generations, SM4,
prone to be the first popular model of new physics on which the LHC will
speak a final verdict.

Within the three generation SM (SM3) the production cross section $\sigma(gg\to
H)$, which governs $pp\to H$ studied at the LHC, is dominated by a triangle
diagram with a top quark. While the loop diagram decreases as $1/m_t$ for $m_t
\to \infty$, this decrease is compensated by the linear growth of the top Yukawa
coupling $y_t \propto m_t$. Consequently, in the SM4 the new contributions from
the heavy $t^\prime$ and $b^\prime$ quarks will modify $\sigma(gg\to H)$ by a
term which is independent of $m_{t^\prime}$ and $m_{b^\prime}$ at the one-loop
level. One finds an increase by roughly a factor of 9, which seemingly entails a
corresponding increase in the LHC signal cross section of Higgs decays into
(virtual) gauge bosons, given by the product $\sigma(pp\to H)\,\Bcal(H\to
WW^*,ZZ^*,\gamma\gamma)$. However, higher-order corrections to the Higgs
production cross sections and branching ratios due to the fourth-generation
fermions can be substantial because of their large Yukawa couplings. In
\cite{Djouadi:1994gf, Djouadi:1994ge, Passarino:2011kv, Denner:2011vt} it was
shown that, for light Higgs bosons, the $H\to WW^*$ and $H\to ZZ^*$ branching
ratios in the SM4 can be suppressed by a factor of $0.2$ or less as compared to
their SM3 values. In the photonic Higgs decay rate $\Gamma(H\to \gamma\gamma)$
the destructive interference between fermion and gauge boson mediated
contributions even leads to an accidental cancellation which would render the
$H\to\gamma\gamma$ decay unobservable. As pointed out in \cite{Djouadi:2012ae},
this leads to tensions with the observed excesses in $H\to\gamma\gamma$ searches
at LHC and the searches for $H\to b\bar b$ in $HW$, $HZ$ associated production
at the Tevatron.

In \cite{Khoze:2001ug, Belotsky:2002ym, Bulanov:2003ka, Rozanov:2010xi,
  Keung:2011zc, Cetin:2011fp, Englert:2011us, Carpenter:2011wb} it was discussed
that the SM4 may permit the decay mode $H\to \nu_4\ov\nu_4$, where $\nu_4$
denotes the neutrino of the fourth generation. If the $\nu_4$ is sufficiently
long-lived, LHC triggers will not associate the $\nu_4$ decay with the primary
Higgs production and decay event, such that $H\to \nu_4\ov\nu_4$ will stay
undetected. That is, with present experimental techniques the mere effect of an
open $H\to \nu_4\ov\nu_4$ channel will be an increase of the total Higgs width
and thus a decrease of all other branching fractions.  In this paper we will
only consider the case of Dirac neutrinos. The fourth-generation neutrino must
therefore be heavier than $M_Z/2$ to comply with the invisible $Z$ width
measured at LEP1. While a nonzero $H\to \nu_4\ov\nu_4$ decay rate can reconcile
the LHC data on $\sigma(pp\to H)\,\Bcal(H\to WW^*,ZZ^*)$ with the SM4, it will
only increase the tensions with the excesses in $H\to\gamma\gamma$ at the LHC
and $H\to b\bar b$ at the Tevatron.

In \cite{Djouadi:2012ae} it was shown that the signals for $H\to\gamma\gamma$
and $H\to b\bar b$ can single-handedly rule out the SM4 if the currently
measured signal cross sections are confirmed with significantly smaller errors.
However, with the current uncertainties one must resort to a global fit to all
relevant observables to assess the viability of the SM4.  The non-decoupling
property of the SM4 implies that the SM3 can not be considered as a special case
of the SM4 where some parameters are fixed. This actually represents a
conceptual problem for a standard frequentist analysis as the choice of a
suitable test statistic for the definition of $p$-values is no longer
straightforward. We do not attempt to solve this issue here. Instead we simply
compare the $\chi^2$ values of the two models and the pulls of the individual
observables. In all our fits we assume that the observed excesses in
$H\to\gamma\gamma$ and $H\to b\bar b$ searches are not statistical fluctuations
and we therefore fix the Higgs mass at $m_H=125\;\text{GeV}$.

Stringent constraints on the SM4 are also found from analyses of the electroweak
precision observables \cite{Nakamura:2010zzi}, because the extra fermions induce
non-decoupling contributions to the $W$ mass, partial $Z$ decay widths and
asymmetries which are very sensitive to the mass splittings within the fermionic
isospin doublets. It has been shown in Ref.~\cite{He:2001tp,
    Novikov:2002tk, Frere:2004rh, Novikov:2009kc, Kribs:2007nz,
    Eberhardt:2010bm} that the SM4 is compatible with the experimental
constraints from LEP if the $m_{t^\prime}$--$m_{b^\prime}$ and/or
$m_{l_4}$--$m_{\nu_4}$ mass splittings are chosen properly. Here $l_4$ denotes
the charged lepton of the fourth generation. In this letter we perform a global
fit to the parameters of the SM4, using the LHC data on the abovementioned Higgs
decays, Tevatron data on $H\to b \ov b$ and electroweak precision data. We also
discuss the impact of mixing between the third and fourth-generation quarks as
well as the impact of an increased lower bound on the fourth generation charged
lepton mass. For our fits we use the CKMfitter package, which implements the
Rfit procedure \cite{Hocker:2001xe}, a frequentist statistical method.
%
%
%
\section{Methodology} 
%
%
The main topic of this letter is a combined fit of the following
(pseudo-)observables, which defines our analysis A1: 
\begin{itemize}
\item[i)] the signal strengths $\hat\mu(pp\to H\to WW^*)$ measured by CMS
  \cite{CMS-PAS-HIG-12-008} (defined below) and $\hat\mu(pp\to H\to ZZ^*)$
  measured by CMS \cite{CMS-PAS-HIG-12-008} and ATLAS \cite{ATLAS:2012ac},
\item[ii)] the signal strengths $\hat\mu(VV\to H\to\gamma\gamma)$ and
  $\hat\mu(gg\to H\to\gamma\gamma)$ for Higgs production via vector boson fusion
  and gluon fusion, respectively, and subsequent decay into two photons as
  measured by CMS \cite{CMS-PAS-HIG-12-001} and ATLAS \cite{ATLAS:2012ad},
\item[iii)] the signal strength $\hat\mu(p\bar p\to HV\to Vb\bar b)$ for Higgs
  production in association with a vector boson and subsequent decay into
  a $b\bar b$ pair, as measured by CDF and D0 \cite{FisherMoriond},
\item[iv)] the electroweak precision observables (EWPOs) $M_Z$, $\Gamma_Z$,
  $\sigma_\text{had}$, $A_\text{FB}^l$, $A_\text{FB}^c$, $A_\text{FB}^b$, $A_l$,
  $A_c$, $A_b$, $R_l=\Gamma_{l^+l^-}/\Gamma_\text{had}$, $R_c$, $R_b$,
  $\sin^2\theta_l^\text{eff}$ measured at LEP and SLC \cite{EWWG:2010vi} as well
  as $m_t$, $M_W$, $\Gamma_W$ and $\Delta\alpha_\text{had}^{(5)}$
  \cite{Nakamura:2010zzi}.
\item[v)] the lower bounds $m_{t^\prime,b^\prime}\gtrsim 600\;$\gev
  (from the LHC) \cite{Aad:2012xc, Aad:2012us, CMS:2012ye, CMS-PAS-EXO-11-099}
  and $m_{l_4}>101\,\text{GeV}$ (from LEP2) \cite{Nakamura:2010zzi}.
\end{itemize}
Here and in the following, the term ``signal strength'' refers to the ratio of
SM4 and SM3 signal cross sections evaluated with the same Higgs mass
\begin{equation}\label{eq:signalstrength}
  \hat\mu(X\to H\to Y) = \frac{\sigma(X\to H)\Bcal(H\to Y)|_\text{SM4}}
                              {\sigma(X\to H)\Bcal(H\to Y)|_\text{SM3}}  
  \quad.
\end{equation}
where a signal cross section is given by the product of the Higgs production
cross section and a branching fraction into a certain final state.

When confronting the SM4 with electroweak precision data, the usual method is to
compute the oblique electroweak parameters $S$ and $T$ \cite{Peskin:1991sw}, and
compare the results to the best-fit values for $S$ and $T$ provided by the LEP
Electroweak Working Group \cite{EWWG:2010vi}. For the SM4, such studies were
done, for example, in Refs.~\cite{Nakamura:2010zzi, Kribs:2007nz, Erler:2010sk,
  Eberhardt:2010bm}. However, it is well-known that the parametrisation of the
EWPOs (iv) by $S$ and $T$ becomes inaccurate when some of the fourth-generation
fermion masses are close to $M_Z$ or when the fourth-generation fermions mix
with the fermions of the first three generations. Since here, we are interested
in a scenario where $m_{\nu_4}<M_Z$ we do not use the oblique electroweak
parameters in our analysis, but fit the EWPOs directly. To this end, we use
ZFitter \cite{Bardin:1989tq, Bardin:1999yd, Arbuzov:2005ma} to compute accurate
predictions for the EWPOs in the SM3. (More precisely, we use the DIZET
subroutine of the ZFitter package.) Then we follow the procedure of
\cite{Gonzalez:2011he} and add corrections due to the fourth-generation fermions
to the EWPOs. The differences between EWPOs in the SM4 and SM3 are calculated at
one-loop order, but no further approximations are made for the EWPOs. As
experimental inputs we use $M_W=80.390\pm 0.016\,\text{GeV}$
\cite{Aaltonen:2012bp} and otherwise the same inputs as the GFitter
collaboration \cite{Baak:2011ze}. With our program we reproduce the best-fit
parameters and observables for the SM3 within less than $10\%$ of the (fit)
error quoted in \cite{Baak:2011ze} for each parameter or observable. Our
electroweak fit differs from the one in \cite{Baak:2011ze} in two points: we
neglect the bottom and charm mass in the calculation of the EWPOs and we do not
include theoretical errors. For the present analysis we also fix the Higgs mass
to $125\,\text{GeV}$.

The current limit on the $b'$ mass according to \cite{CMS:2012ye} is
approximately $600\;\text{GeV}$. However, this and other limits on fourth
generation quark masses by CMS and ATLAS rely on certain assumptions about the
decay pattern of the quarks. These limits can be severely weakened if CKM mixing
and `cascade decays' such as $t'\to b'W$ are taken into account
\cite{Flacco:2010rg}.  In this letter we avoid the bounds on fourth-generation
quark masses by fixing the $b^\prime$ mass to $m_{b^\prime}=600\,\gev$. The
splitting between the fourth-generation quark masses is strongly constrained by
the EWPOs, so that the bound on $m_{t'}$ will automatically be satisfied.

In close correspondence to SM3 electroweak fits such as \cite{Nakamura:2010zzi,
  Baak:2011ze}, we let the following parameters float in our fit:
\begin{equation}
\Delta\alpha_\text{had}^{(5)},\;  \alpha_{s},\;  M_Z, \;  m_t,\; 
m_{t^\prime}, \;  m_{\nu_4},\;  m_{l_4}\;  \mbox{and}\; 
\theta_{34},
\label{pars}
\end{equation}
where $\Delta\alpha_\text{had}^{(5)}$ is the hadronic contribution to the
running of the fine-structure constant in the 5-flavour scheme and $\theta_{34}$
denotes the mixing angle between the third and fourth generation, defined
analogously to the Cabibbo angle. The importance of the mixing angle
$\theta_{34}$ in the SM4 electroweak fit was pointed out in
\cite{Chanowitz:2009mz}.  Mixing of the fourth generation with the first two
generations and additional $CP$ violating phases can be relevant if flavour
observables are included in the fit. However, the constraints on these
  parameters from flavour physics are so strong that the allowed
  variations do not have a big effect on the observables studied in this
  letter. We therefore set these additional phases and mixing angles to zero.
Note that we fix the Higgs mass to 125$\;$\gev, which is the value favoured by
the hints seen in 2011 LHC data. The choice of a fixed value for $m_{b^\prime}$
does not lead to a significant loss of generality, as the experimental lower
bound $m_{b^\prime}\gtrsim 600\;\text{GeV}$ \cite{CMS:2012ye} is
already rather close to the scale where the Yukawa interactions become
non-perturbative \cite{Chanowitz:1978uj}. Also, the non-decoupling property of
the most relevant quantities implies a rather mild dependence on $m_{b^\prime}$.

We include the two-loop electroweak corrections to Higgs production and decay in
our evaluation of the Higgs signal cross sections in the SM4 by means of the
program HDECAY v.\ 4.45 \cite{Djouadi:1997yw}. This is mandatory, because the
flat dependence of these decay amplitudes on $m_{t^\prime,b^\prime,l_4}$ is
broken by the leading two-loop corrections \cite{Denner:2011vt}. To avoid the
complicated procedure of interfacing the HDECAY code with our program we set ---
for the evaluation of the Higgs signal cross sections ---
$m_{t'}=650\,\text{GeV}$, $\theta_{34}=0$ and the SM parameters $\alpha$,
$\alpha_{s}$, $M_Z$ and $m_t$ to the default values of HDECAY. The dependence of
the cross sections on $m_{\nu_4}$ and $m_{l_4}$ is then accounted for by linear
interpolation of two-dimensional lookup-tables with a granularity of
$0.5\,\text{GeV}$ for $m_{\nu_4}$ and $50\,\text{GeV}$ for $m_{l_4}$. As the
experimental errors on the Higgs signal cross sections are still rather large
this simplification has no noticable impact on our fit.

Table \ref{tab:signalstrengths} summarises our experimental inputs for the Higgs
signal strengths in the different search channels:
\begin{table}
  \centering
  \renewcommand{\arraystretch}{1.3}
  \begin{tabular}{lll}
    \hline\hline
    process & signal strength & reference \\
    \hline
    $VV\to H\to\gamma\gamma$ & $3.7^{+2.0}_{-1.7}$    &
    \cite{CMS-PAS-HIG-12-001} \\
    $gg\to H\to\gamma\gamma$ & $1.30^{+0.49}_{-0.50}$ &
    \cite{CMS-PAS-HIG-12-001, ATLAS:2012ad} \\
    $pp\to H\to WW^*$        & $0.39^{+0.61}_{-0.56}$ &
    \cite{CMS-PAS-HIG-12-008} \\
    $pp\to H\to ZZ^*$        & $0.69^{+0.93}_{-0.52}$ &
    \cite{CMS-PAS-HIG-12-008, ATLAS:2012ac} \\
    $p\bar p\to HV\to Vb\bar b$ & $2.03^{+0.73}_{-0.71}$ &
    \cite{FisherMoriond} \\
    \hline\hline
  \end{tabular}
  \caption{Experimental inputs for Higgs signal strengths at $m_H=125\;$\gev.}
  \label{tab:signalstrengths}
\end{table}
The signal strength for Higgs production via vector boson fusion (VBF) and
subsequent decay into $\gamma\gamma$ ($VV\to H\to\gamma\gamma$) corresponds to
the signal strength for the dijet class in \cite{CMS-PAS-HIG-12-001}.  We assume
that the events in this category stem entirely from vector boson fusion
processes. This is, of course, a somewhat crude approximation. There will also
be a certain contamination from gluon fusion events in that sample, but lacking
more detailed information on this contamination we are forced to ignore it.  The
signal strength for Higgs production via gluon fusion and subsequent decay into
$\gamma\gamma$ ($gg\to H\to\gamma\gamma$) was obtained by removing the dijet
contribution from the combined result for the signal strength in
\cite{CMS-PAS-HIG-12-001} and combining the result with the one from
\cite{ATLAS:2012ad}. In doing this, we implicitly neglect all Higgs production
mechanisms except gluon fusion and vector boson fusion.  The signal strength for
$pp\to H\to ZZ^*$ is a combination of the results presented in
\cite{CMS-PAS-HIG-12-008} and \cite{ATLAS:2012ac}. The signal strength for
$pp\to H\to WW^*$ was taken from \cite{CMS-PAS-HIG-12-008}. The input for the
$p\bar p\to HV\to Vb\bar b$ process is taken from the latest Tevatron search
\cite{FisherMoriond} for Higgs bosons produced in association with a $W$ or $Z$
boson and subsequently decaying into a $b\bar b$ pair.

For the computation of signal cross sections in the SM4 we use an effective
coupling approximation along the lines of \cite{Bechtle:2008jh,
  Bechtle:2011sb}. Specifically, we calculate the SM4 signal cross sections by
taking SM3 production cross sections for the different production mechanisms
from \cite{Dittmaier:2011ti} (LHC) and \cite{Brein:2003wg, Baglio:2010um}
(Tevatron), scaling them with corresponding SM4/SM3 ratios of related partial
Higgs decay widths and multiplying with the SM4 branching fractions calculated
by HDECAY. For instance, the SM4 signal cross section for $gg\to
H\to\gamma\gamma$ is calculated as
\begin{multline}\label{eq:effc}
  \sigma(gg\to H\to\gamma\gamma)_\text{SM4}
  =\sigma(gg\to H)_\text{SM3}\\
   \times\frac{\Gamma(H\to gg)_\text{SM4}}{\Gamma(H\to gg)_\text{SM3}}
   \Bcal(H\to\gamma\gamma)_\text{SM4}\ ,
\end{multline}
with $\sigma(gg\to H)_\text{SM3}$ taken from \cite{Dittmaier:2011ti} and the
remaining quantities on the right-hand side calculated by HDECAY.  The factor
$\Gamma(H\to gg)_\text{SM4}/\Gamma(H\to gg)_\text{SM3}$ accounts for the
modified $Hgg$ effective coupling in the SM4. For the VBF process $VV\to
H\to\gamma\gamma$ the Higgs can come from a $HWW$ or $HZZ$ vertex. We assume
that $75\%$ of the production cross section comes from $WW$ fusion and $25\%$
from $ZZ$ fusion. These ratios were obtained from \cite{SpiraVV2H}, which
implements the NLO results from \cite{Han:1992hr}. Equations analogous to
\eqref{eq:effc} are then used separately for the $WW\to H$ and $ZZ\to H$
production modes. For the $pp\to H\to WW^*$ and $pp\to H\to ZZ^*$ signal cross
sections all production mechanisms were taken into account.  For the (Tevatron)
$p\bar p\to HV\to Vb\bar b$ process only the $HW$ and $HZ$ associated production
mechanisms contribute. The corresponding SM3 production cross sections were
taken from \cite{Baglio:2010um}. Like the LHC cross sections these were scaled
with the SM4/SM3 ratios of $H\to WW$ and $H\to ZZ$ partial widths, respectively,
and multiplied with the SM4 $H\to b\bar b$ branching fraction.

In order to disentangle the impacts of the Higgs searches and the electroweak
precision observables we perform a second fit, denoted as analysis A2. In this
analysis we only fit the Higgs data, ignoring the EWPOs altogether. Here we only
let $m_{\nu_4}$ and $m_{l_4}$ float, while keeping $m_{t^\prime}$ fixed to
$650\,\text{GeV}$.
%
%
%
\section{Results}
%
%

From Table~\ref{tab:signalstrengths} we see that the searches for $VV\to
H\to\gamma\gamma$ and $p\bar p\to HV\to Vb\bar b$ prefer an enhancement of the
SM signal while the searches for $pp\to H\to WW^*$ and $pp\to H\to ZZ^*$ prefer
reduced signals. Thus, only the $pp\to H\to WW^*$ and $pp\to H\to ZZ^*$ searches
favour a large invisible Higgs decay width and our fits must choose a neutrino
mass that compromises between the two tendencies. The result of our analysis A2
(fitting Higgs signal strengths only) is $m_{\nu_4}=60\;\text{GeV}$ and
$m_{l_4}=600\;\text{GeV}$ (the latter being the upper end of the range in which
$m_{l_4}$ was allowed to float). So, the best-fit neutrino mass is just below
the $H\to\nu_4\bar\nu_4$ threshold, leading to $\Bcal(H\to\nu_4\bar\nu_4)\approx
0.46$.  The minimum $\chi^2$ value in this fit is $18.2$. This should
be compared to the $\chi^2$ value of 6.1, which is obtained in the
SM3. These results agree with a recent analysis of this type by Kuflik, Nir and
Volansky \cite{Kuflik:2012ai}. Their conversion of the $\chi^2$ values to
confidence levels should however be taken with a grain of salt since, in their
analysis, some of the SM4 parameters were scanned over but not counted as
degrees of freedom when converting $\chi^2$ values into confidence levels.  In
general, the number of degrees of freedom of a fit is ill-defined when
parameters are only allowed to float within a certain range (such as the fourth
generation fermion masses) and the relation between the $\chi^2$ value and the
confidence level is no longer described by the normalised lower incomplete gamma
function. Due to the afore-mentioned conceptual problems with the definition of
a suitable test statistic for the comparison of SM4 and SM3 we refrain from
converting our $\chi^2$ values into $p$-values and only discuss the pulls of the
individual signal strengths. We hope to shed more light on the issue of a
quantitative comparison of the SM3 and SM4 in a future publication.

The best-fit charged lepton mass in the analysis A2 is at the
upper end of the range in which it was allowed to float. Of course, such
a large mass splitting within the lepton doublet is ruled out by
electroweak precision data. In our analysis A1 (combination of EWPOs and
Higgs signal strengths) we obtain the following best-fit values:
\begin{gather}
  m_{\nu_4}=58.0\;\gev
  \ ,\qquad  
  m_{l_4}=113.6\;\gev
  \ ,\nonumber\\
  m_{t'}=632\;\gev
  \ ,\qquad
  \chi^2_\text{SM4,min} = 33.4
  \ .
\end{gather}
We see that the best-fit charged lepton mass is now just above the LEP limit.
The best-fit neutrino mass has moved to a slightly lower value, leading to
$\Bcal(H\to\nu_4\bar\nu_4)\approx 0.66$. The minimum $\chi^2$ value
should be compared with the SM3 value $\chi^2_\text{SM3,min} = 21.7$.

\begin{figure}
  \includegraphics[width=0.36\textwidth,bb=160 494 440 737,clip=true]{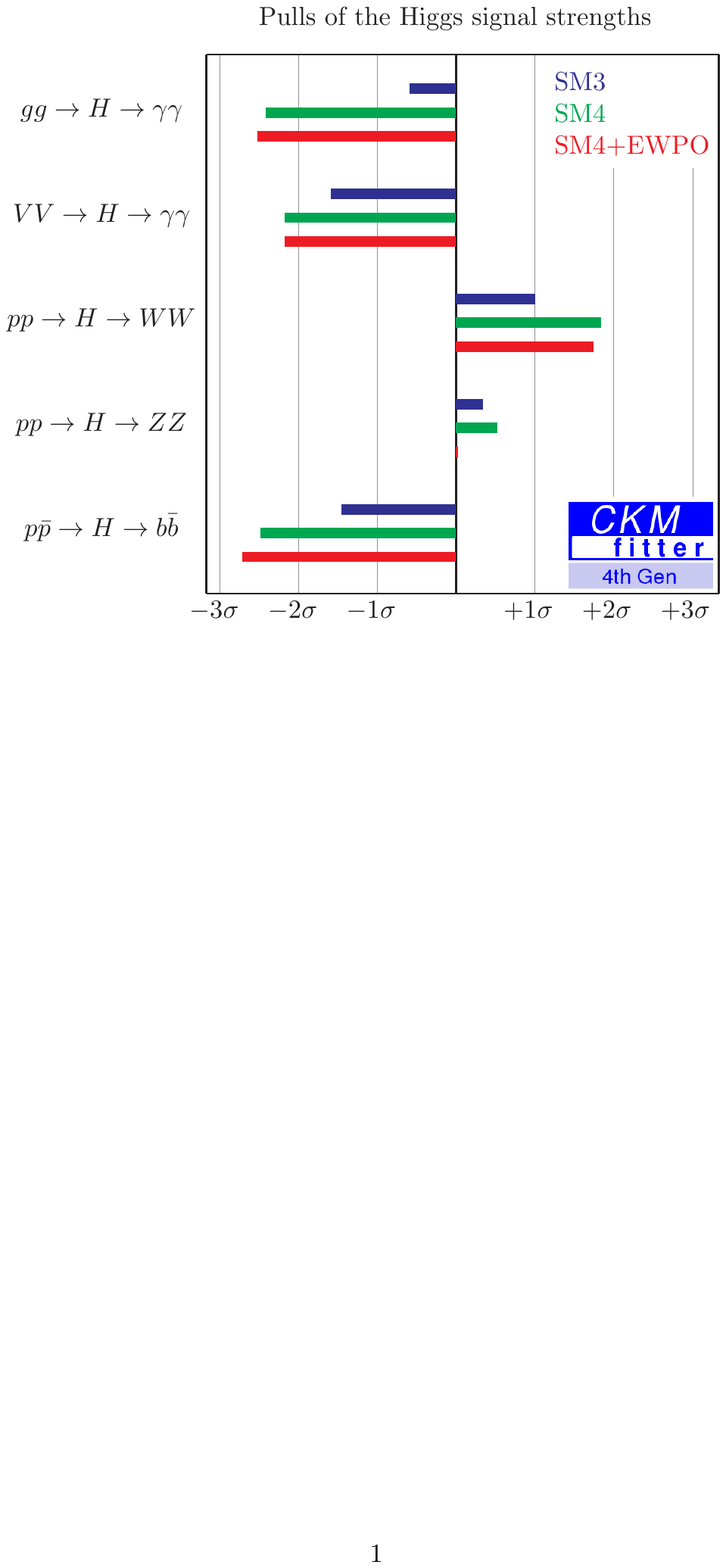}
  \caption{Deviations (pulls) of the predicted signal strengths from the
    measured signal strengths in units of the experimental errors. The pulls are
    shown for the SM3 and the two SM4 scenarios, corresponding to our analyses
    A1 (SM4 w.~EWPO) and A2 (SM4 w/o EWPO).}
  \label{fig:pulls1}
\end{figure}
Figure~\ref{fig:pulls1} shows the pulls of the signal strengths in the SM3 and
SM4 for our analyses A1 and A2. The pulls are defined as
$(\hat\mu_\text{pred}-\hat\mu_\text{exp})/\Delta \hat\mu$, where
$\hat\mu_\text{exp}$ and $\Delta \hat\mu$ are the experimental values and errors
of the signal strengths in Table~\ref{tab:signalstrengths} and
$\hat\mu_\text{pred}$ is obtained by removing the experimental input for the
corresponding signal strength from the fit and using the other observables to
predict its value. We see that the pulls for the analyses A1 and A2 are
essentially the same. This can be understood as follows: the main effect of
including the EWPOs in the fit is that the lepton mass is constrained to smaller
values, but the Higgs signal strengths are not sensitive enough to the lepton
mass for this to make a big difference.  With the exception of $pp\to H\to
ZZ^*$, the pulls in the SM4 are always bigger than in the SM3, their magnitude
being around $2\sigma$. For $pp\to H\to ZZ^*$ the predicted SM4 signal strength
is equal to the measured one while the pull in the SM3 is about
$0.5\sigma$. This agreement of the SM4 is however purely accidental.

\begin{figure}
  \includegraphics[width=0.4\textwidth,bb=133 477 484 713,clip=true]{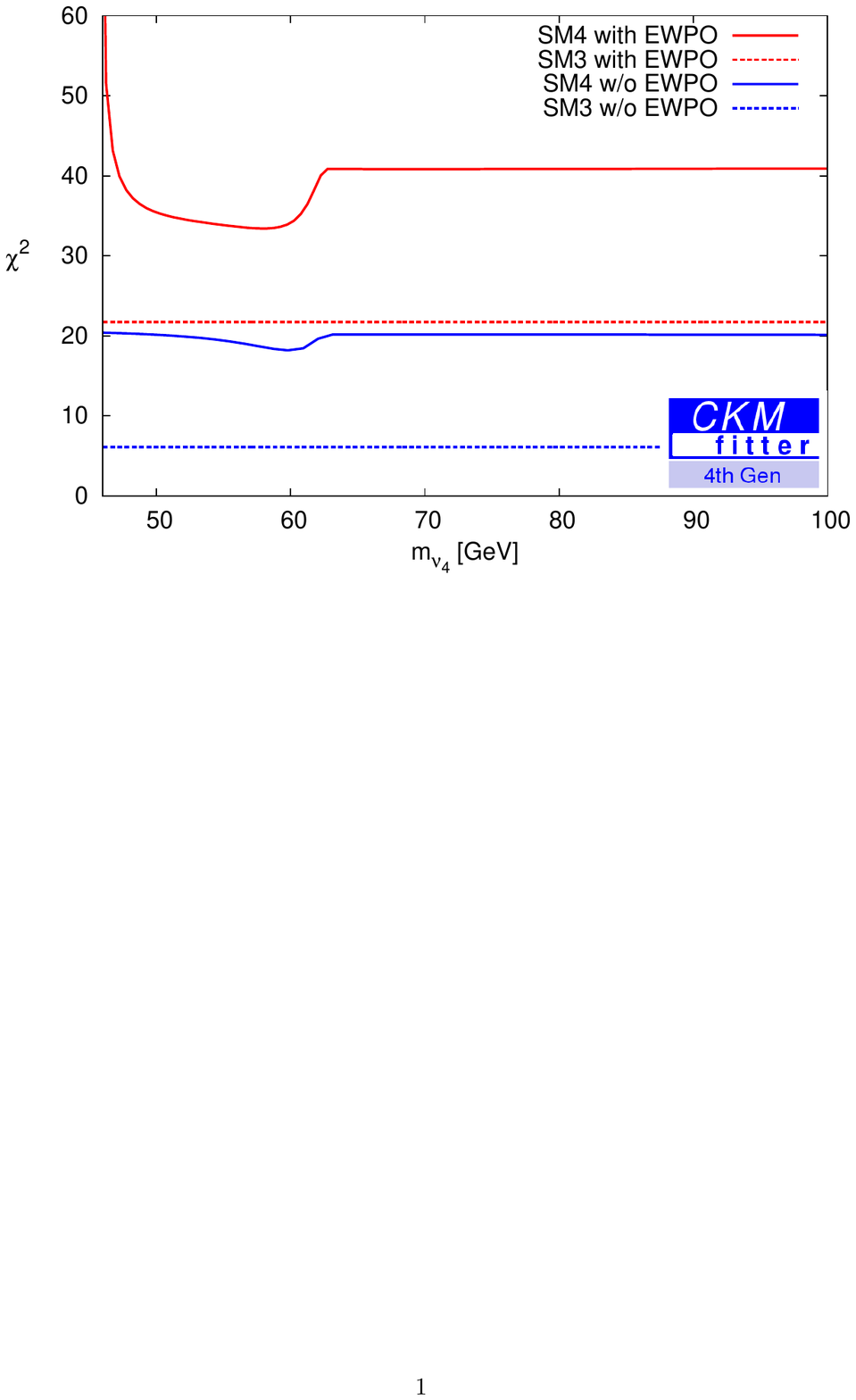}
  \caption{Minimum $\chi^2$ values for a fixed neutrino mass as a function of
    $m_{\nu_4}$. Results are shown for the two SM4 analyses A1 (red) and A2
    (blue). The dotted lines indicate the corresponding SM3 minimal $\chi^2$
    value.}
  \label{fig:chi1}
\end{figure}
In \fig{fig:chi1} we show the minimum $\chi^2$ as a function of $m_{\nu_4}$ and
minimised with respect to the other parameters in \eqref{pars} for our analyses
A1 and A2. The $\chi^2$ value of the SM3 is indicated by the dotted line.  We
see that the SM3 has a smaller $\chi^2$ value than the SM4 for any choice of
$m_{\nu_4}$. In both analyses the best-fit value of $m_{\nu_4}$ is near
$60\;\text{GeV}$, i.e.\ just below the $H\to \nu_4\bar\nu_4$ threshold.  For
$m_{\nu_4}\lesssim 60\;\text{GeV}$ the Higgs signal strengths favour a small
lepton mass while for $m_{\nu_4}\gtrsim 60\;\text{GeV}$ a large charged lepton
mass is preferred by direct Higgs searches. Since EWPOs forbid too large mass
splittings (of order $100\;\text{GeV}$ or more) in the lepton doublet the
increase of $\chi^2$ at $m_{\nu_4}\approx 60\;\text{GeV}$ is more pronounced in
the analysis A1.  Above the $H\to\nu_4\bar\nu_4$ threshold the $\chi^2$ value is
essentially independent of $m_{\nu_4}$. As $m_{\nu_4}$ approaches $M_Z/2$ the
$\chi^2$ in the analysis A1 blows up due to threshold effects in the EWPOs.

\begin{figure}
  \includegraphics[width=0.4\textwidth,bb=132 476 484 715,clip=true]{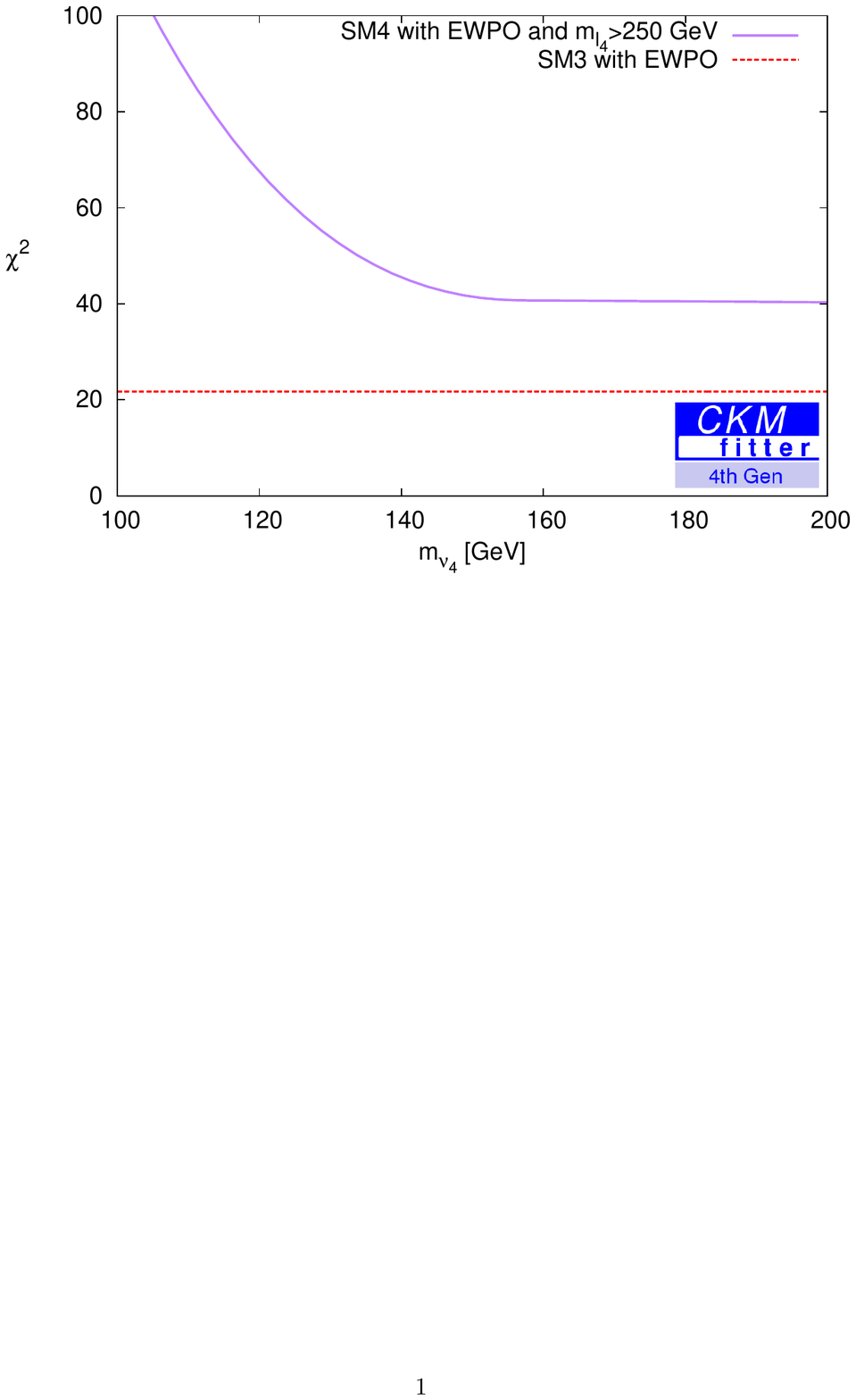}
  \caption{Minimum $\chi^2$ values in the analysis A1 for
    $m_{l_4}>250\;\text{GeV}$ as a function of the (fixed) neutrino mass
    $m_{\nu_4}$. The dotted lines indicate the corresponding SM3 minimal
    $\chi^2$ value.}
  \label{fig:chi2}
\end{figure}

In a sensitivity study \cite{Carpenter:2010bs} for fourth-generation charged
lepton searches at the LHC it was found that with $1\;\text{fb}^{-1}$ of data
the LHC experiments should be able to rule out a fourth-generation charged
lepton with a mass below approximately $250\;\text{GeV}$. Currently there are no
experimental results available for these searches. Let us nonetheless
investigate what happens if the mass bound for the fourth-generation charge
lepton moves up to $250\;\text{GeV}$. Fig.~\ref{fig:chi2} shows the $\chi^2$ of
our analysis A1 with a modified charged lepton mass limit
$m_{l_4}>250\;\text{GeV}$ as a function of $m_{\nu_4}$, minimized with respect
to all other parameters.  We see that the $\chi^2$ is constant at a value of
$36$ for $m_{\nu_4}\gtrsim 160\;\text{GeV}$.  For neutrino masses below
$160\;\text{GeV}$ the electroweak fit can no longer accomodate the large mass
splitting in the lepton sector and the $\chi^2$ blows up. Thus, for
$m_{l_4}>250\;\text{GeV}$ (and the case of Dirac neutrinos) the scenario with
the invisible $H\to\nu_4\bar\nu_4$ decay is completely ruled out by electroweak
precision observables.

\begin{figure}
  \includegraphics[width=0.4\textwidth,bb=133 477 477 712,clip=true]{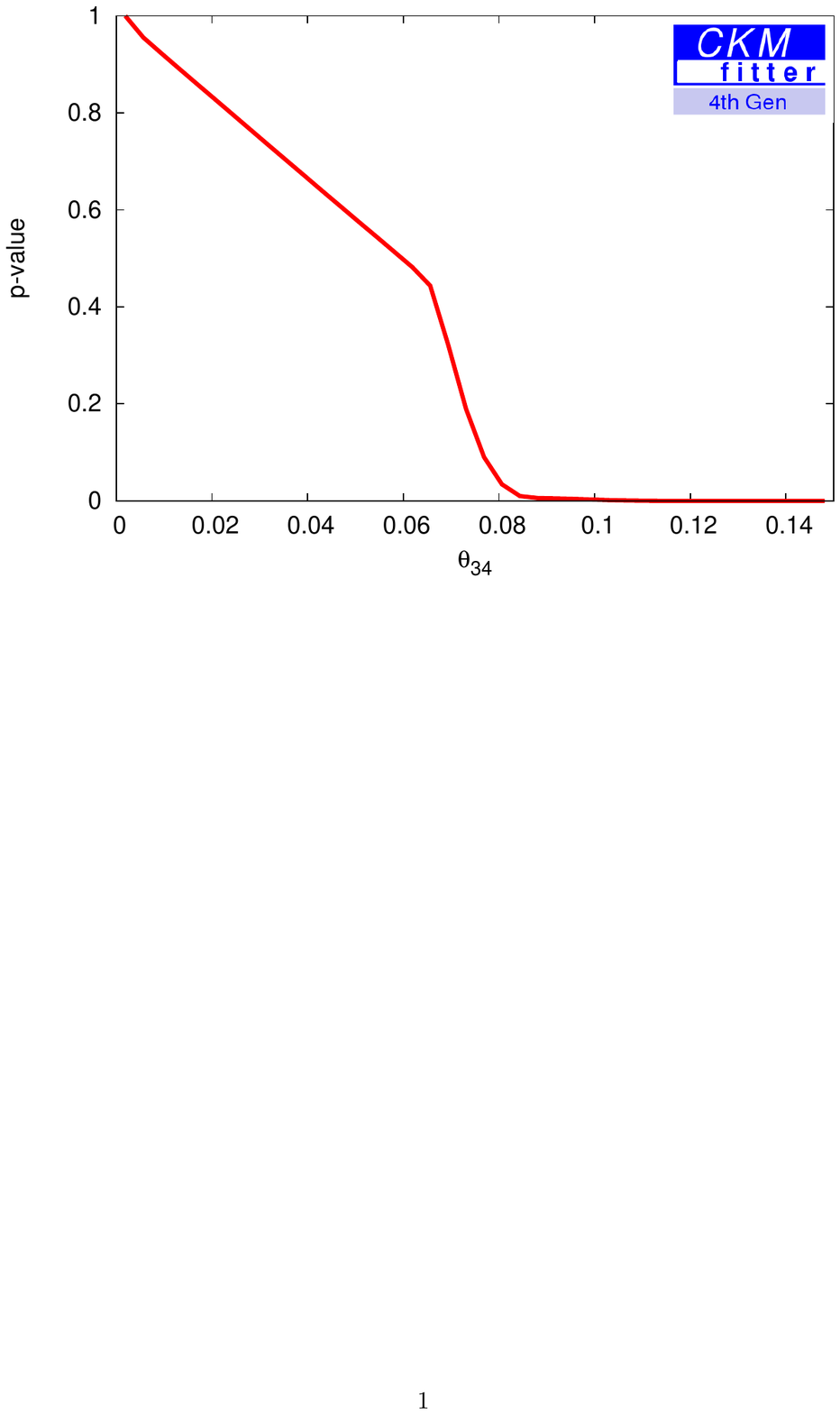}
  \caption{$p$-value scan of the CKM mixing angle $\theta_{34}$ between the
    third and fourth generation quarks in the analysis A1.}
  \label{fig:theta34}
\end{figure}

The impact of mixing between the third and fourth generation quark is negligible
in the analysis A1. The fit prefers $\theta_{34}=0$ and therefore cannot be
improved by letting $\theta_{34}$ float. The constraint on $\theta_{34}$ imposed
by EWPOs and Higgs signal strengths can be studied by using the difference
between the minimal $\chi^2$ in the SM4 with $\theta_{34}$ free and
$\theta_{34}$ fixed as a test statistic. Since we are now comparing two
different realisations of the same model (SM4) there is no problem with the
conversion of $\chi^2$ values to $p$-values. Fig.~\ref{fig:theta34} shows the
$p$-value as a fuction $\theta_{34}$. We see that Higgs signal strengths and
EWPOs require $\theta_{34}\lesssim 0.08$. However, this picture could change
dramatically if flavour observables were included in the fit: A recent analysis
shows that the SM3 fails to describe flavour physics observables at the level of
$2.7\sigma$ \cite{Lenz:2010gu, Lenz:2012az, Bevan:2010gi, Lunghi:2010gv}. Since
the SM4 can alleviate the discrepancies in the flavour data, the overall picture
may still change in favour of the SM4 in a complete analysis of Higgs decay,
electroweak precision, and flavour data. Such an analysis is beyond the scope of
this letter.
%
%
%
\section{Conclusions}
%
%

Assuming a Higgs mass of $125\;\text{GeV}$ we have performed a global fit to the
parameters of the SM4, combining data on electroweak precision physics and five
different Higgs searches: $H\to\gamma\gamma$ produced by gluon fusion at the
LHC, $H\to\gamma\gamma$ produced by vector boson fusion at the LHC, inclusive
searches for $H\to WW,ZZ$ at the LHC and $W$,$Z$ associated production and decay
to $b\bar b$ at the Tevatron.  With the exception of the inclusive $H\to ZZ$
search the pulls of the signal cross sections in the SM4 exceed those of the SM3
by $0.5\sigma$ or more. Also the electroweak precision observables are described
better in the SM3. With a lower bound of $100\;\text{GeV}$ on the
fourth-generation charged lepton mass the best-fit SM4 scenario has a
fourth-generation neutrino mass around $60\;\text{GeV}$, i.e.\ just below the
$H\to\nu_4\bar\nu_4$ threshold. If the lower bound on the fourth-generation
charged lepton mass moves up to $250\;\text{GeV}$ the electroweak precision
observables constrain $m_{\nu_4}$ to be larger than approximately
$160\;\text{GeV}$ and scenarios with invisible $H\to\nu_4\bar\nu_4$ decays are
ruled out.  The mixing angle $\theta_{34}$ between the third and fourth
generation quarks is constrained to be smaller than $0.08$.  However, since the
SM4 can alleviate the discrepancies in flavour observables, the overall picture
may still change in favour of the SM4 when flavour observables are included in
the fit. On the basis of electroweak precision data and Higgs searches alone the
SM4 is certainly disfavoured. A quantitative comparison of the SM3 and SM4 in
terms of $p$-values is problematic since classical likelihood ratio tests for
nested models are inapplicable due to the non-decoupling nature of the SM4
fermions.  We hope to shed more light on this subject in a future publication.

\acknowledgments

We would like to thank the CKMfitter group, in particular J\'er{\^o}me Charles
and St\'ephane T'Jampens, for valuable input on statistical methods and
CKMfitter software support. Further, we would like to thank Julien Baglio for
fruitful discussions and help with the VBF cross sections. We acknowledge
support by DFG through grants NI1105/2-1 and LA 2541/1-1. A.L.\ is further
supported by DFG through a Heisenberg fellowship.

\bibliography{sm4higgs}

\end{document}